\documentclass[aps,prl,twocolumn,superscriptaddress,showpacs]{revtex4}

\usepackage{graphicx}% Include figure files
\usepackage{dcolumn} % Align table columns on decimal point
\usepackage{bm}      % bold math
\DeclareGraphicsExtensions{.pdf,.gif,.eps}

\newcommand{\Tn}{{T$_N$}}
\newcommand{\U}{{URu$_2$Si$_2$}}
\newcommand{\usr}{{$\mu$SR}}

\newcommand{\ub}{{$\mu_{B}$}}

\begin{document}

\preprint{APS/123-QED}

\title{Hidden orbital currents and spin gap in the heavy fermion superconductor {\U}}

\author{C.~R.~Wiebe}
\email{wiebecr@mcmaster.ca}
\affiliation{Department of Physics and
Astronomy, McMaster University, Hamilton, Ontario L8S 4M1, Canada}
\affiliation{Department of Physics, Columbia University, New York,
New York 10027, USA}

\author{G.~M.~Luke}
\affiliation{Department of Physics and Astronomy, McMaster
University, Hamilton, Ontario L8S 4M1, Canada}
\author{Z.~Yamani}
\affiliation{Department of Physics, University of Toronto,
Toronto, Ontario M5S 1A7, Canada}
\author{A. A. Menovsky}
\affiliation{Van der Waals - Zeeman Institute, University of
Amsterdam, Valckenierstraat 65, 1018 XE Amsterdam, The
Netherlands}
\author{W.~J.~L.~Buyers}
\affiliation{Neutron Program for Materials Research, NRC, Chalk
River Laboratories, Chalk River, Ontario K0J 1J0, Canada}

\date{\today}% It is always \today, today,

\begin{abstract}

We have performed neutron scattering experiments on the heavy
fermion superconductor {\U} to search for the orbital currents
predicted to exist in the ordered phase below {\Tn} = 17.5 K.
Elastic scans in the (H, K, 0) and (H, 0, L) planes revealed no
such order parameter at low temperatures.  This does not
completely rule out orbital current formation, because our
detection limit for a ring of scattering is 0.06(1) {\ub}, which
is greater than the size of the predicted moment of ~0.02 {\ub}.
However, on heating, a ring of quasielastic scattering does exist
in the (H, K, 0) plane centered at the (1, 0, 0) antiferromagnetic
Bragg position and of incommensurate radius {$\tau$} = 0.4 r.l.u..
The intensity of this ring is thermally activated below {\Tn} with
a characteristic energy scale of {$\Delta$} = 110 K: the coherence
temperature.  We believe that these incommensurate spin
fluctuations compete with the AF spin fluctuations, and drive the
transition to a disordered magnetic state above {\Tn}. The
significance of this higher energy scale with respect to {\Tn}
suggests that these fluctuations also play a crucial role in the
formation of the heavy fermion state.

\end{abstract}

%\keywords{muon spin relaxation}%Use showkeys class option if keyword

\maketitle

We have examined the prediction of Chandra $\emph{et al.}$ that
orbital currents may exist in strongly correlated electron systems
\cite{Chandra}.  High temperature superconductivity,
superfluidity, the quantum Hall Effect, and heavy fermion metals
are all examples of the fascinating behavior that arises from
complicated interactions within systems of fermions
\cite{Fisk},\cite{Heffner}. The interest in the latter example,
heavy fermion metals, has been piqued due to the discovery of the
coexistence of superconductivity and antiferromagnetism in several
species such as UPt{$_3$}\cite{Aeppli},
UPd{$_2$}Al{$_3$},\cite{Geibel} UNi{$_2$}Al{$_3$},\cite{Lussier}
and {\U} \cite{Broholm}.  A striking feature of these compounds is
that they display conduction electron specific heats at low
temperatures that are orders of magnitude greater than found in
typical metals. This behavior differs markedly from the high
temperature behavior, at which the value for the Sommerfeld
constant is returned to normal free electron values. As evidenced
from the resistivity, specific heat, and magnetic susceptibility,
these compounds are well described at high temperatures as a set
of weakly interacting conduction electrons and local moments.  The
crossover from this state to a low temperature state, in which the
effective mass of the quasiparticles increases dramatically, is a
gradual change that is characterized by a coherence temperature
$\theta$$_{C}$.\cite{Amato}

The physics of the heavy fermion state is often posed as a
competition between the Kondo effect, in which conduction
electrons screen out local magnetic moments, and the RKKY
interaction, which polarizes the conduction electrons and provides
an oscillatory exchange constant as a function of distance between
magnetic sites, thus enhancing local moment
formation.\cite{Amato}, \cite{Doniach} The magnitudes of both
effects increase as a function of the density of quasiparticle
states at the Fermi energy g(E{$_f$}), but they have different
functional forms.  At high values of g(E{$_f$}), the Kondo
interaction is stronger than the RKKY, and thus some heavy
fermions do not order at all such as CeAl{$_3$} \cite{Andres}. But
for low values of g(E{$_f$}), there is a tendency for local moment
formation, albeit of reduced size (such as UNi{$_2$}Al{$_3$} and
UPd{$_2$}Al{$_3$}, which have ordered U{$^{4+}$} moments of 0.24
{\ub} \cite{Lussier} and 0.85 {\ub} \cite{Krimmel} respectively).

In the case of UPt{$_3$} and {\U}, the ordered moments in the
N{\'{e}}el states are extremely small (0.02 {\ub} \cite{Aeppli2}
and 0.03 {\ub} \cite{Broholm} respectively).  UPt{$_3$} has an
exotic superconducting state involving multiple transitions as a
function of applied pressure and magnetic fields \cite{Willis}.
The scenario for {\U} is intriguing as well: the transition at
{\Tn} = 17.5 K is accompanied by a large lambda anomaly in the
specific heat, \cite{Palstra} which, in light of the extremely
small ordered moment, suggests that another order parameter is at
play.  There has been considerable interest in {\U} over the last
few years, with new developments providing hints that the ordered
magnetic state is inhomogeneous,\cite{Luke} and somewhat parasitic
to a so-called ``hidden'' ordered state.  A complex phase diagram
for this state has been elucidated based upon specific heat
measurements at high magnetic fields.\cite{Kim} The character of
this possible hidden order parameter is still unknown, but
possibilities such as quadrupolar ordering and charge-density wave
formation are still plausible.\cite{Santini}  It has been shown,
however, that none of the allowed quadrupolar or octupolar
orderings can account for the weak moment.\cite{Walker}  Recent
NMR measurements have shown that small isotropic magnetic fields
develop below {\Tn} at the silicon sites,\cite{Matsuda},
\cite{Bernal} which have led Chandra $\emph{et al.}$ to develop a
theory of orbital current formation \cite{Chandra},
\cite{Chandra2}. The signature for such currents which develop in
the ordered phase below {\Tn} would be a ring of incommensurate
scattering in reciprocal space with a characteristic Q{$^{-4}$}
form factor.

We have carried out a detailed search for this hidden order
parameter in {\U} using neutron scattering in the (H, K, 0) and
(H, 0, L) planes. No such scattering is found to exist within
experimental error. However, we have found an unusual ring of
quasielastic scattering at an incommensurate radius from the (1,
0, 0) antiferromagnetic zone center which decreases in spectral
weight below {\Tn}. The spectral weight shifts from this ring of
scattering, whose characteristic energy scale is the coherence
temperature, to the antiferromagnetic spin correlation wavevector
and suggests that competing interactions play a role in the
ordering at {\Tn}.

Our experiments were performed at the DUALSPEC triple-axis
spectrometer at Chalk River Laboratories with pyrolytic graphite
as monochromator and analyzer set to a fixed energy of 3.52 THz. A
graphite filter was used to remove higher order contamination.
Elastic scans were performed, as well as quasielastic scans at an
energy transfer of 0.25 THz. The collimation was chosen to be
0.40$^{o}$-0.48$^{o}$-0.56$^{o}$-1.20$^{o}$.  The crystals were
the same two used earlier \cite{Broholm}, one oriented in the (H,
K, 0) plane and the other in the (H, 0, L) plane.

Figure 1 shows the (1, 0, 0) magnetic Bragg signal arising from
the 0.03 {\ub} ordered moment. The relative intensity of this peak
as compared to background gives a measure of our sensitivity to
ordered moments.  Considering the signal to background ratio, our
calculations indicate that the minimum size of a detectable signal
for long-ranged local moment formation is 0.013(1) {\ub} for 3D
order (resulting in a 3D Bragg peak).  For a moment distribution
which leads to a 2D ring structure with the 0.2 r.l.u. radius of
Chandra \cite{Chandra}, this limit changes to 0.06(1) {\ub} (one
can calculate this by evaluating the ratios of the areas of a ring
structure compared to a magnetic Bragg peak in Q-space,
considering our instrumental resolution).

\begin{figure}[t]
 \linespread{1}
 \begin{center}
  \includegraphics[scale=0.4,angle=0]{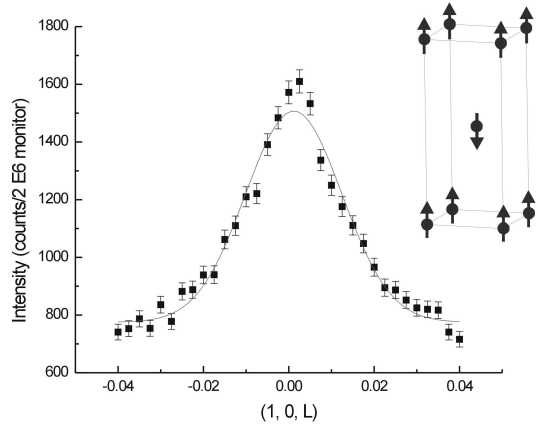}
  \caption{The antiferromagnetic Bragg peak at (1, 0, 0) and T = 8 K (fit is to a Gaussian).  The inset shows the corresponding magnetic structure for the U$^{4+}$ moments.}
  \label{lattice}
 \end{center}
 \linespread{1.6}
\end{figure}

\begin{figure}[t]
 \linespread{1}
 \begin{center}
  \includegraphics[scale=0.4,angle=0]{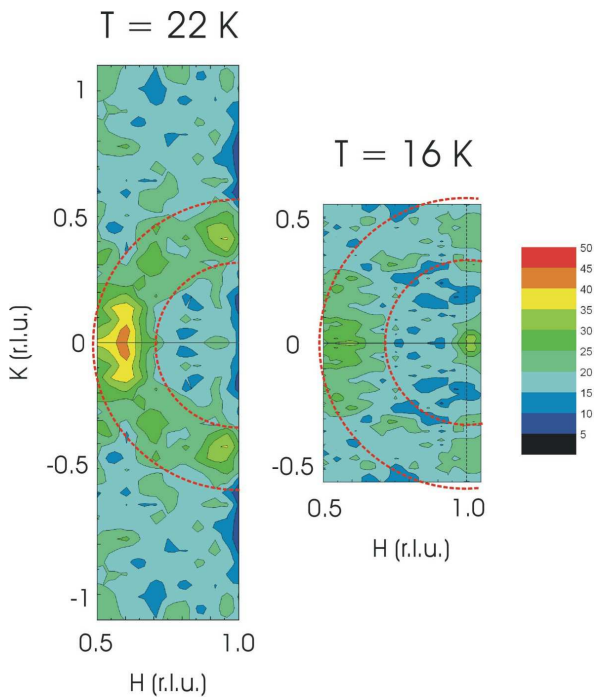}
  \caption{(Color online) Contour plot of scattering in the (H, K, 0) plane at T = 22 K  and T = 16 K with $\Delta$E = 0.25 THz energy transfer.  Note the ring of
  scattering centered about (1, 0, 0), and the antiferromagnetic fluctuations at (1, 0, 0) associated with the order parameter.  The dashed circles are guides to the eye.}
  \label{ring22k}
 \end{center}
 \linespread{1.6}
\end{figure}

Elastic raster scans were made to look for hidden order with
ranges 0.5 $\leq$ H $\leq$ 1.0 and 0 $\leq$ K $\leq$ 1.05 for the
(H, K, 0) plane and 0.175 $\leq$ H $\leq$ 1.075 and 0 $\leq$ L
$\leq$ 1.05 for the (H, 0, L) plane.  The difference scans, I(8 K)
- I(22 K), indicate that there are no new sources of magnetic
Bragg scattering below {\Tn}.  This, together with the results of
Bull $\emph{et al.}$, indicate that no new magnetic Bragg peaks
exist in the (H, H, L), (H, K, 0), and (H, 0, L) planes
\cite{Bull}.  However, this does not completely rule out orbital
current formation, because our detection limit for a ring of
scattering is 0.06(1) {\ub}, which is greater than the size of the
moment predicted by Chandra \emph{et al} of ~0.03 {\ub}.
\cite{Chandra}, \cite{Chandra2}  Note that for body-centered
tetragonal symmetry, the position for the predicted ring of
scattering at ({$\tau$} cos{$\theta$}, {$\tau$} sin{$\theta$}, 1)
\cite{Chandra}, \cite{Chandra2} is equivalent to the wavevectors
(1 + {$\tau$} cos{$\theta$}, {$\tau$} sin{$\theta$}, 0) at which
we made the search.

Since no new features were discovered in the elastic channel, we
decided to look at the quasielastic spectra at $\Delta$E = 0.25
THz. This removes the large incoherent elastic peak and so
increases the sensitivity to the formation of slow correlations
modulated in Q. Our strategy was to further investigate an
incommensurate ring of scattering which was discovered in the
previous investigation of the inelastic spectrum near (1.4, 0, 0).
A broad feature centered at about 0.6 THz was reported above
{\Tn}, indicative of heavily damped antiferromagnetic spin
fluctuations.\cite{Broholm} This feature sharpened to a nearly
resolution limited peak in energy below {\Tn} with a center at
higher energies ($\sim$ 1.1 THz). This results in an increase in
scattering as one passes above {\Tn}.  The structure of this
scattering in reciprocal space at this energy transfer was not
reported in the original paper, nor was its explicit temperature
dependance.\cite{Broholm} Subsequent work suggested that the
structure could be a ring in reciprocal space, \cite{Broholm2},
\cite{Mason} which is what Chandra $\emph{et al}$ predicted for
orbital current formation from the hidden order phase.

Figure 2 shows contours of the scattering in the (H, K, 0) plane
above and below T$_{N}$ at 22 K and 16 K.  We have folded the data
about the line K = 0 because of the symmetry of reciprocal space.
 The scans at T = 22 K show a ring of scattering centered in the (H,
K, 0) plane at AF zone centers such as (1, 0, 0) and of radius
{$\tau$} = 0.4 r.l.u.. These are not powder lines, which would be
centered at Q = 0, but they are antiferromagnetic spin
fluctuations centered about the AF points such as (1, 0, 0) and
(2, 1, 0). Note the absence of this scattering about (1, 1, 0), a
ferromagnetic point. To confirm that these features are magnetic
in origin, the form factor has been measured out to higher values
of Q.  Figure 3 shows the results of these scans, in which points
at (0.6, 0, 0), (1.4, 0, 0), (2, 0.6, 0) and (2, 1.4, 0) were
measured.  The U{$^{4+}$} magnetic form factor, plotted on the
same curve, is in good agreement with our data.  The scattering
does not exhibit the predicted Q$^{-4}$ form factor that Chandra
$\emph{et al.}$ predicted for orbital currents, although it does
form a magnetic ring in reciprocal space.

\begin{figure}[t]
 \linespread{1}
 \begin{center}
  \includegraphics[scale=0.4,angle=0]{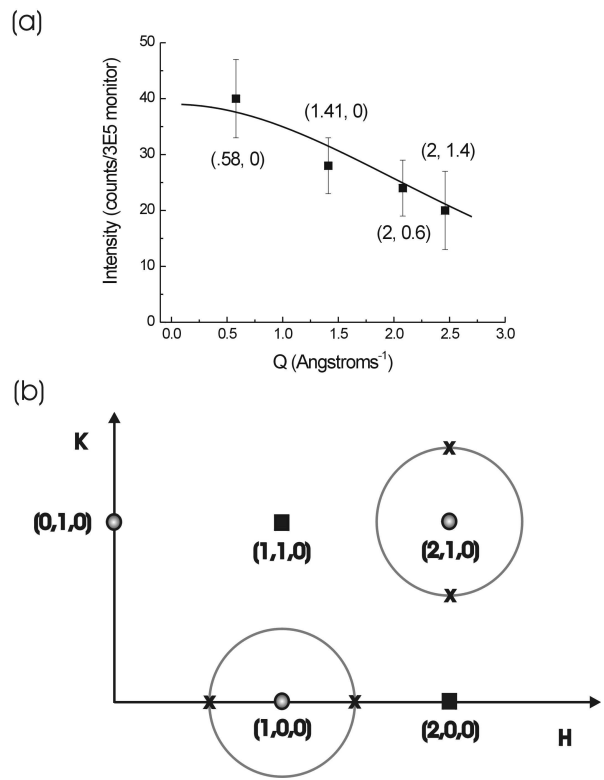}
  \caption{(a)  The background subtracted intensities (as determined by Gaussian fits) to the features seen at rings of scattering in the (H, K, 0) plane.  The line is the U$^{4+}$ magnetic form factor \cite{Brown}.
  (b)  A map of the rings in reciprocal space with crosses where the form factor was measured.  The squares and circles refer to ferromagnetic and antiferromagnetic points, respectively.}
  \label{kspacemap}
 \end{center}
 \linespread{1.6}
\end{figure}

The ring of scattering is thermally activated up to the transition
at {\Tn} as shown in figure 4. This qualitatively agrees with the
previous work of Broholm \textit{et al.} \cite{Broholm} who
measured the inelastic spectra above and below the transition.
What we are observing is the tail of an inelastic peak above
{\Tn}, which moves outside our energy window as the peak narrows
and moves to a higher energy scale below {\Tn} (resulting in a
suppression of intensity below the transition).

The fit in figure 4 is to a background plus a single activated
intensity of the form:

\begin{equation}
I(T)= A\exp(-{\Delta}/{T})~~ T < T_{N}
\end{equation}
\begin{equation}
I(T)= {\rm{constant}} ~~ T > T_{N}
\end{equation}

where A is a constant, {\Tn} is the N\'{e}el temperature (17.5 K),
and $\Delta$ = 110(10) K is the fitted activation energy.  It is
important to note the activation temperature is not that of the
sampling energy, 0.25 THz $\sim$ 12 K, nor that of the 0.67 THz
$\sim$ 30 K spin excitation above {\Tn}.  Instead it corresponds
to that found by Palstra \textit{et el.} for the specific heat
anomaly below {\Tn}.\cite{Palstra}  In the range of (2 K - 17.5
K), the specific heat could be fit to the following:

\begin{eqnarray}
{C(T) = {\gamma}T + {\beta}T^{3} + {\delta}\exp{^{-{\Delta}/T}}}
\label{Palstra}
\end{eqnarray}

with $\Delta$ $\sim$ 115 K.  This suggested that a substantial gap
opens in the density of states below {\Tn} (which when interpreted
as a momentum space phenomenon, gaps $\sim$ 75 \% of the Fermi
surface).\cite{Palstra}  This gap is similar to the maximum in the
spin-wave density of states,\cite{MasonBuyers} and the gap seen in
the optical reflectance measurements.\cite{Bonn} The existence of
two competing energy scales ($\Delta$ and {\Tn}) has also been
observed in DC resistivity experiments.\cite{Mentink}

\begin{figure}[t]
 \linespread{1}
 \begin{center}
  \includegraphics[scale=0.45,angle=0]{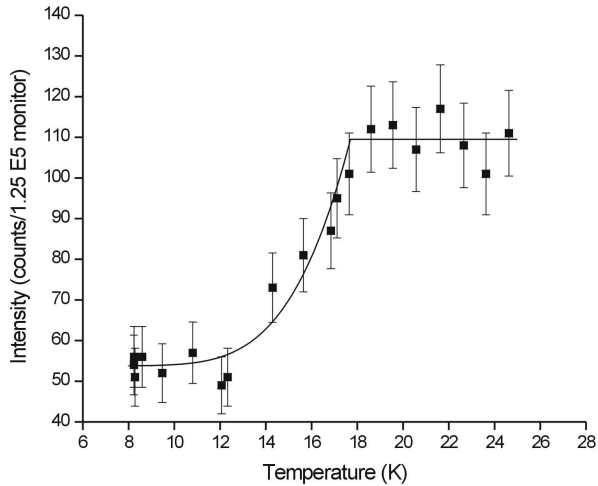}
  \caption{The temperature dependance of the scattering at (1.4, 0, 0) and $\Delta$E = 0.25 THz energy transfer.  The fit is with an activation temperature of 110(10) K}
  \label{tempdep}
 \end{center}
 \linespread{1.6}
\end{figure}

The quasielastic scattering observed in this experiment is
reminiscent of the intimate connection between spin fluctuations
and the heavy fermion state.  Gaulin \textit{et al.} have studied
this relationship in UNi{$_2$}Al{$_3$}, which has a {\Tn} of 4.6 K
and T$_{c}$ of 1.2 K.\cite{Gaulin},\cite{Schroeder}  The magnetic
structure is incommensurate, with an ordering wavevector of Q =
(0.5 $\pm$ $\tau$, 0, 0.5) ($\tau$ = 0.11) and an ordered moment
of 0.85 {\ub} per U atom. The quasielastic spin fluctuations are
of two kinds: those associated with the incommensurate wavevector,
and with the commensurate Q = (0, 0, 0.5) wavevector.  The two
modes compete with one another, with a shift in spectral weight
from the commensurate to incommensurate fluctuations below {\Tn}.
Above the transition, the incommensurate fluctuations disappear,
but the commensurate ones persist to nearly 80 K, the coherence
temperature. This is strong evidence that these excitations are
associated with the formation of the heavy fermion state. It is
remarkable that both {\U} and UNi{$_2$}Al{$_3$} show the
competition between two excitations associated with different
energy scales ({\Tn} and $\Delta$).  In the case of
UNi{$_2$}Al{$_3$}, the commensurate fluctuations were noted to be
at the same ordering wavevector as in the sister compound
UPd{$_2$}Al{$_3$}, which orders at {\Tn} = 14.5 K.  A competition
between the RKKY interaction and the Kondo effect is believed to
explain the difference in the magnetic structures and spin
excitation spectra of UNi{$_2$}Al{$_3$} and UPd{$_2$}Al{$_3$}.
\cite{Gaulin}

For {\U}, the situation is reversed:  the commensurate
fluctuations are associated with the ordering wavevector (1, 0,
0), and the incommensurate excitations persist to high
temperatures (and therefore can be identified with the formation
of the heavy fermion state). This shift in spectral weight can be
noted by the increase in scattering at (1, 0, 0) below \Tn~ (see
figure 2) and the corresponding decrease in intensity at (1.4, 0,
0). Mason {\em et al.} have suggested that magnetic frustration
plays a role in the unusual magnetic properties of {\U}, with the
long range oscillatory nature of the RKKY interaction providing
the mechanism.\cite{Mason}  The incommensurate ring of scattering
we observe may be a signature of such a RKKY interaction. However,
the true nature of these fluctuations cannot be ascertained with
this experiment alone. The fact that the activation energy is the
same for the spin response and the specific heat anomaly strongly
suggests that these are excitations out of the gapped ``hidden
order'' state rather than the AF ordered phase.  Further
experiments are needed to develop a clearer picture of the origin
of these excitations with respect to proposed scenarios of a phase
separation occurring at {\Tn}.  It may be that this feature is
linked electronic phase separation, as suggested by {\usr}
\cite{Luke} and NMR measurements,\cite{Matsuda} since the spectral
weight is too large to be explained by the 0.03 {\ub} ordered
moment.

In conclusion, our neutron scattering measurements in search of
hidden order in the (H, K, 0) and (H, 0, L) planes have placed an
upper limit of 0.013(1) {\ub} for the presence of any long-ranged
3D ordered spin structure well-defined in Q.  For a ring of
scattering, the detectable moment is 0.06(1) {\ub}, which
precludes orbital current formation that is somewhat larger than
that predicted by Chandra $\emph{et al.}$. \cite{Chandra}
Quasielastic scattering experiments have revealed a connection
between a ring of incommensurate scattering at a radius of
{$\tau$} = 0.4 from the zone center and the heavy-fermion state.
The exponential activation energy of this ring below {\Tn},
comparable with the specific heat activation energy, suggests that
a gap of 110 K is a feature of the ``hidden order'' phase.

\begin{acknowledgments}

C.~R.~Wiebe would like to acknowledge support from NSERC.  The
authors would like to thank the financial support of NSERC and
CIAR. Useful discussions with B.~D.~Gaulin and C.~Broholm were
appreciated. The authors are also grateful for the technical
support of the NPMR staff at Chalk River.

\end{acknowledgments}

\end{document}